\def\year{2020}\relax
\documentclass[letterpaper]{article} 
\usepackage{aaai20}  
\usepackage{times}  
\usepackage{helvet} 
\usepackage{courier}  
\usepackage[hyphens]{url}  
\usepackage{graphicx} 
\usepackage{algorithm}
\usepackage{subfigure}
\usepackage[noend]{algpseudocode}
\usepackage{algorithmicx,algorithm}
\usepackage{booktabs}
\usepackage{latexsym}
\usepackage{amsmath}
\usepackage{comment}
\usepackage{amsfonts}
\usepackage{xcolor}
\usepackage{multirow}
\usepackage{verbatim}
\usepackage{multicol}
\usepackage{url}
\usepackage{array}
\usepackage{color}
\usepackage{graphicx}  
\title{When Low Resource NLP Meets Unsupervised Language Model:   \\Meta-pretraining Then Meta-learning for Few-shot Text Classification}

\author{
Shumin Deng\textsuperscript{1,2}    \thanks{\quad All authors contributed equally to this work.}
Ningyu Zhang\textsuperscript{2,4}    \footnotemark[1]
   Zhanlin Sun\textsuperscript{2,5}   \footnotemark[1]
 Jiaoyan  Chen\textsuperscript{6}   
 Huajun Chen\textsuperscript{1,2,3}\thanks{\quad Corresponding author.}\\
1. College of Computer Science and Technology, Zhejiang University\\
 2. Alibaba-Zhejiang University Frontier Technology Research
 Center Joint Lab for Knowledge Engine\\
 3. The First Affiliated Hospital of Zhejiang University\\
 4. Alibaba Group \\
 5. School of Computer Science, Carnegie Mellon University\\
 6. Department of Computer Science, Oxford University\\
  \tt \{231sm,huajunsir\}@zju.edu.cn, zhanlins@andrew.cmu.edu \\
  \tt jiaoyan.chen@cs.ox.ac.uk, ningyu.zny@alibaba-inc.com}
 
\begin{document}

\maketitle

\begin{abstract}
Text classification tends to be difficult when data are deficient or when it is required to adapt to unseen classes. In such challenging scenarios, recent studies have often used meta-learning to simulate the few-shot task, thus negating implicit common linguistic features across tasks. This paper addresses such problems using meta-learning and unsupervised language models. Our approach is based on the insight that having a good generalization from a few examples relies on both a generic model initialization and an effective strategy for adapting this model to newly arising tasks.  We show that our approach is not only simple but also produces a state-of-the-art performance on a well-studied sentiment classification dataset. It can thus be further suggested that pretraining could be a promising solution for few-shot learning of many other NLP tasks.  The code and the dataset to replicate the experiments are made available at \url{https://github.com/zxlzr/FewShotNLP}.
\end{abstract}

\section{Introduction}
Deep learning (DL) has achieved great success in many fields owing to the advancements in optimization techniques, large datasets, and streamlined designs of deep neural architectures. However, DL is notorious for requiring large labeled datasets, which limits the scalability of a deep model to new classes owing to the cost of annotation.  Few-shot learning generally resolves the data deficiency problem by recognizing novel classes from very few labeled examples. This limitation in the size of samples (only one or very few examples) challenges the standard fine-tuning method in DL. Early studies in this field applied data augmentation and regularization techniques to alleviate the overfitting problem caused by data scarcity but only to a limited extent. Instead, researchers have been inspired by exploration of meta-learning \cite{finn2017model} to leverage the distribution over similar tasks.  However, existing meta-learning approaches for few-shot learning can not explicitly disentangle task-agnostic and task-specific representations, and they are not able to take advantage of the knowledge of linguistic properties via unsupervised language models. 

In this paper, we raise the question that \textbf{whether it is possible to boost the performance of low-resource natural language processing with the large scale of raw corpus via unsupervised leaning}, which require us to handle both task-agnostic and task-specific representation learning.   Thus we propose a \textbf{M}eta-pretraining \textbf{T}hen \textbf{M}eta-learning  (\textbf{MTM}) approach motivated by the observation that meta-learning leads to learning a better parameter initialization for new tasks than multi-task learning across all tasks.  The former meta-pretraining is to learn task-agnostic representations that explicitly learns a model parameter initialization for enhanced predictive performance with limited supervision. The latter meta-learning considers all classes as coming from a joint distribution and seeks to learn model parameters that can be quickly adapted via using each class's training instances to enhance predictive performance on its test set. In other words, our approach explicitly disentangles the task-agnostic and task-specific  feature learning. Experimental results   demonstrate that the proposed model achieves significant improvement  on public benchmark datasets.   

\section{Approach}

\subsection{Problem Definition}
Few-shot text  classification \cite{yu2018diverse,geng2019few} is a task in which a classifier must adapt new classes that are not seen in training, given only a few examples for each of these new classes. To be specific, we have a labeled training set with a set of defined classes $\mathcal{C}_{train}$. Our goal is to output classifiers on the testing set with a disjoint set of new classes $\mathcal{C}_{test}$ when only a small labeled support set is available. If the support set contains $K$ labeled examples for each of the $C$ unique classes, the target few-shot problem is called a $C$-way-$K$-shot problem. The sample set  is usually too small to train a supervised classification model. To this end, we try to utilize meta-learning method on the training set  to extract task-agnostic knowledge, which may   perform better for  few-shot text classification on  the test set.

\subsection{Training Procedure}

\textbf{Task-agnostic Meta Pretraining.} Given all the training samples, we first utilize pretraining strategies such as BERT to learn task-agnostic contextualized features that capture linguistic properties to benefit downstream
few-shot text classification tasks. 

\textbf{Meta-learning Text Classification.} Given the pretrained language representations, we construct episodes to compute gradients and update the model in each training iteration. 

\begin{algorithm}[th] 
\begin{algorithmic}[1]
\caption{MTM Algorithm} 
\Require   Training Datapoints  $\mathcal{D}=\left\{\mathbf{x}^{(j)}, \mathbf{y}^{(j)}\right\}$
 \State  Construct a task $T_j$ with training examples using a support set  $\mathcal{S}_{K}^{(j)}$  and a test example $\mathcal{D}_{j}^{\prime}=\left(\mathbf{x}^{(j)}, \mathbf{y}^{(j)}\right)$
\State Randomly initialize $\theta$
\State Pre-train  $\mathcal{D}$ with  unsupervised   language models 
\State Denote $p(\mathcal{T})$ as distribution over tasks
\While{not done}
\State Sample batch of tasks  $\mathcal{T}_{i} \sim p(\mathcal{T})$:
\For {for all $T_i$}
 \State  Evaluate   $\nabla_{\theta} \mathcal{L}_{\mathcal{T}_{i}}\left(f_{\theta}\right)$ using  $\mathcal{S}_{K}^{(j)}$
\State Compute adapted parameters with gradient descent: $\theta_{i}^{\prime}=\theta-\alpha \nabla_{\theta} \mathcal{L}_{\mathcal{T}_{i}}\left(f_{\theta}\right)$
\EndFor
\State  Update $\theta \leftarrow \theta-\beta \nabla_{\theta} \sum_{\mathcal{T}_{i} \sim p(\mathcal{T})} \mathcal{L}_{\mathcal{T}_{i}}\left(f_{\theta_{i}^{\prime}}\right)$
using each $\mathcal{D}_{i}^{\prime}$ from $T_{i}$ and  $\mathcal{L}_{\mathcal{T}_{i}}$
\EndWhile
\label{alg} 
\end{algorithmic}
\end{algorithm}

\section{Experiments}
 \subsection{Datasets and Evaluation}

We use the multiple tasks with the multi-domain sentiment classification  dataset ARSC\footnote{https://github.com/Gorov/DiverseFewShot\_Amazon}. This dataset comprises English reviews for $23$ types of products on Amazon. For each product domain, there are three different binary classification tasks. These buckets then form $23 \times 3 = 69$ tasks in total. We select $12 (4 \times 3)$ tasks from four domains  as the test set, with only five examples as support set for each label in the test set.  We evaluate the performance by few-shot classification accuracy following previous studies in few-shot learning \cite{snell2017prototypical}. To evaluate the proposed model objectively with the baselines, note that for ARSC, the support set for testing is fixed by \cite{yu2018diverse}; therefore, we need to run the test episode once for each of the target tasks. The mean accuracy from the $12$ target tasks are compared to those of the baseline models in accordance with \cite{yu2018diverse}.

\subsection{Evaluation Results}
 The evaluation results are shown in Table \ref{exp1}:   
\textbf{MTM} is our current approach, \textbf{Match Network} \cite{vinyals2016matching} is a few-shot learning model using metric-based attention method, \textbf{Prototypical Network} \cite{snell2017prototypical} is a deep matrix-based method using sample averages as class prototypes,  \textbf{MAML} \cite{finn2017model} is a model-agnostic method that is compatible with any model trained with gradient descent and applicable to a variety of learning problems, \textbf{Relation Network} \cite{sung2018learning} is a metric-based few-shot learning model that uses a neural network as the distance measurement and calculate class vectors by summing sample vectors in the support set, \textbf{ROBUSTTC-FSL} \cite{yu2018diverse} is an approach that combines adaptive metric methods by clustering the tasks, \textbf{Induction-Network-Routing} \cite{geng2019few} is a recent state-of-the-art method which learn generalized class-wise representations by combining the dynamic routing algorithm with a typical meta-learning framework. From the results shown in Table \ref{exp1}, we observe that our approach achieves the best results amongst all meta-learning models.  Note that, our model is task-agnostic, which means it can be easily adapted to any other NLP tasks.

\begin{table}[!htbp]
 
\centering
\begin{tabular}{cc}
\toprule
\textbf{Model}  & \textbf{Mean Acc}\\
\midrule
Matching Network  & 65.73\\ 
Prototypical Network    & 68.15\\ 
Relation Network     & 83.74\\ 
MAML   & 78.33 \\ 
 ROBUSTTC-FSL   & 83.12  \\
 Induction-Network-Routing   & 85.47\\ 
\textbf{MTM} & \textbf{90.01*} \\
\bottomrule
\end{tabular}
\caption{Comparison of mean accuracy (\%) on ARSC. * indicates $p_{value} < 0.01$ in a  paired t-test (10-fold) evaluation.}
\label{exp1}
\end{table}

\section{Conclusion}
In this study, we attempt to analyze language meta-pretraining with meta-learning for few-shot text classification.  
Results show that our model outperforms conventional state-of-the-art few-shot text classification models.  In the future, we plan to apply our method to other NLP scenarios. 

\section{Acknowledgments}
We want to express gratitude to the anonymous reviewers for their hard work and kind comments and this work is funded by NSFC 91846204, national key research program 2018YFB1402800, and Alibaba CangJingGe(Knowledge Engine) Research Plan.

\bibliography{AAA-ZhangN.SA41}
\bibliographystyle{aaai}

\end{document}